\documentstyle[12pt,aaspp4,psfig]{article}

\lefthead{Simpson et al.}
\righthead{The aligned radio galaxy 3C~256}

\begin{document}

\title{Young stars and non-stellar emission in the aligned radio
galaxy 3C~256\footnote{Based in part on observations made with the
NASA/ESA {\it Hubble Space Telescope\/}, obtained at the Space
Telescope Science Institute, which is operated by the Association of
Universities for Research in Astronomy, Inc.\ under NASA contract
NAS~5-26555, and on observations made with the W. M. Keck
Observatory}}

\author{Chris Simpson\altaffilmark{2},
Peter Eisenhardt\altaffilmark{3,4},
Lee Armus\altaffilmark{5,6},
Arati Chokshi\altaffilmark{7},
Mark Dickinson\altaffilmark{8,4},
S. G. Djorgovski\altaffilmark{6},
Richard Elston\altaffilmark{9},
Buell T. Jannuzi\altaffilmark{10},
Patrick J. McCarthy\altaffilmark{11}
Michael A. Pahre\altaffilmark{12}, and
B. T. Soifer\altaffilmark{6,5}}

\altaffiltext{2}{Subaru Telescope, National Astronomical Observatory
of Japan, 650 N.~A`oh\={o}k\={u} Place, Hilo, HI 96720}
\altaffiltext{3}{Jet Propulsion Laboratory, California Institute of
Technology, MS 169--327, 4800 Oak Grove Drive, Pasadena, CA 91109}
\altaffiltext{4}{Visiting observer, Kitt Peak National Observatory,
National Optical Astronomy Observatories, which is operated by AURA,
Inc.\ under cooperative agreement with the National Science
Foundation}
\altaffiltext{5}{SIRTF Science Center, California Institute of
Technology, Pasadena, CA 91125}
\altaffiltext{6}{Palomar Observatory, California Institute of
Technology, Pasadena, CA 91125}
\altaffiltext{7}{Indian Institute of Astrophysics, Bangalore 560 034,
India}
\altaffiltext{8}{Space Telescope Science Institute, 3700 San Martin
Drive, Baltimore, MD 21218}
\altaffiltext{9}{Department of Astronomy, University of Florida, SSRB
211, Gainesville, FL 32611}
\altaffiltext{10}{National Optical Astronomy Observatories, 950
N.~Cherry Avenue, P.O. Box 26732, Tucson, AZ 85726}
\altaffiltext{11}{Observatories of the Carnegie Institute of
Washington, 813 Santa Barbara Street, Pasadena, CA 91101}
\altaffiltext{12}{Harvard-Smithsonian Center for Astrophysics, 60
Garden Street, Mail Stop 20, Cambridge, MA 02138}

\begin{abstract}
We present ground-based images of the $z=1.824$ radio galaxy 3C~256 in
the standard {\it BVRIJHK\/} filters and an interference filter
centered at 8800\,\AA, a {\it Hubble Space Telescope\/} image in a
filter dominated by Ly$\alpha$ emission (F336W), and spectra covering
rest-frame wavelengths from Ly$\alpha$ to [\ion{O}{3}]~$\lambda$5007.
Together with published polarimetry observations, we use these to
decompose the overall spectral energy distribution into nebular
continuum emission, scattered quasar light, and stellar emission.  The
nebular continuum and scattered light together comprise half (one
third) of the $V$-band ($K$-band) light within a 4\arcsec\ aperture,
and are responsible for the strong alignment between the
optical/near-infrared light and the radio emission. The stellar
emission is dominated by a population estimated to be 100--200\,Myr
old (assuming a Salpeter IMF), and formed in a short burst with a peak
star formation rate of 1--$4 \times 10^3\,M_\odot$\,yr$^{-1}$. The
total stellar mass is estimated to be no more than $2 \times
10^{11}\,M_\odot$, which is far less than other luminous radio
galaxies at similar redshifts, and suggests that 3C~256 will undergo
further star formation or mergers.
\end{abstract}

\section{Introduction}

The radio galaxy 3C~256 ($z = 1.824$; Dey et al.\ 1996, hereafter D96)
was first identified by Spinrad \& Djorgovski (1984), at which time it
was the highest redshift galaxy known. It is one of the most luminous
radio galaxies so far discovered ($L_{178MHz} = 3.5 \times
10^{29}$\,W\,Hz$^{-1}$; we adopt $H_0 = 50$\,km\,s$^{-1}$\,Mpc$^{-1}$,
$q_0 = 0.1$, and $\Lambda = 0$\footnote{For this cosmology, the
angular size scale at the distance of 3C~256 is
11.3\,kpc\,arcsec$^{-1}$, and the look-back time is 11.9\,Gyr to a
Universe that is 4.6\,Gyr old.}), and also one of the best examples of
the alignment effect, i.e., the tendency for the optical and radio
emission from high-redshift radio galaxies to share a common position
angle (Chambers, Miley, \& van Breugel 1987; McCarthy et al.\
1987). However, unlike most luminous radio galaxies, where the
alignment is much weaker in the near-infrared (Rigler et al.\ 1992;
Dunlop \& Peacock 1993), in 3C~256 the continuum shows a clear
elongation along the radio position angle of $\sim 140\arcdeg$ even in
the $K$-band (Eisenhardt \& Dickinson 1992). The radio and optical
sizes of 3C~256 are also very similar, with a 4\arcsec\ (45\,kpc)
separation between the two radio lobes.

The nature of the alignment effect is not yet fully understood.
Various mechanisms have been proposed, including star formation
induced by the expanding radio source (e.g., Rees 1989; De Young 1989;
Begelman \& Cioffi 1989), scattering of an anisotropic continuum from
the active nucleus (e.g., di Serego Alighieri et al.\ 1989; Tadhunter
et al.\ 1989; Fabian 1989), inverse Compton scattering of microwave
background photons (Daly 1992a,b) and nebular continuum emission
(e.g., Dickson et al.\ 1995). All of these models explain why the
alignment effect is normally observed to be stronger in the rest-frame
ultraviolet, since the aligned continuum is much bluer than the
evolved stellar population which is believed to emit most of the
near-infrared light in these radio galaxies. Since 3C~256 displays
such a strong alignment at $K$ ($\lambda_{\rm rest} \approx
7800$\,\AA), is unusually blue ($R - K = 2.4$; Eisenhardt \& Dickinson
1992), and is underluminous in the $K$-band Hubble diagram ($K \approx
19$; Chokshi \& Eisenhardt 1991), it may be an object which has yet to
form a substantial population of evolved stars. Such arguments led
Elston (1988) and Eisenhardt \& Dickinson (1992) to suggest that
3C~256 is undergoing its first major burst of star formation, i.e., is
a protogalaxy.

Imaging polarimetry observations by Jannuzi et al.\ (1995) have shown
the spatially extended emission to be strongly polarized ($P = 11.7
\pm 1.5$\% in a 3\farcs6 aperture; $17.6 \pm 2.2$\% in a 7\farcs8
aperture) with the electric vector perpendicular to the radio and
optical axes of 3C~256. This result is naturally explained if the
extended polarized radiation is produced by the scattering of emission
from a source located near the center of 3C~256. Similar polarimetric
observations of other radio galaxies have been used to support the
scattering hypothesis for the alignment effect. Spectropolarimetry by
D96 further supports this scenario in the case of 3C~256 by
demonstrating that the spatially extended polarized radiation has a
power law form consistent with the emission having been originally
produced by an active nucleus ($S_\nu \propto \nu^{-1.1\pm0.1}$).
These authors also suggested a protogalactic nature for 3C~256,
proposing that the galaxy has yet to undergo the major burst(s) of
star formation that will convert most of its mass into stars.

In this paper, we combine new and existing optical and near-infrared
observations to identify and model the components of the spectral
energy distribution of 3C~256. In \S2, we present the new data
obtained for this paper. In \S3, we construct the rest-frame
ultraviolet/optical spectral energy distribution of 3C~256 and discuss
its properties and the salient morphological points. In \S4, we
identify the components necessary to explain the observed SED and
morphology and estimate their strengths. In \S5, we discuss the
implications of our decomposition, and in \S6, we provide a summary of
our results.

\section{Observations and reduction}

\subsection{Optical spectroscopy}

\begin{figure}
\plotfiddle{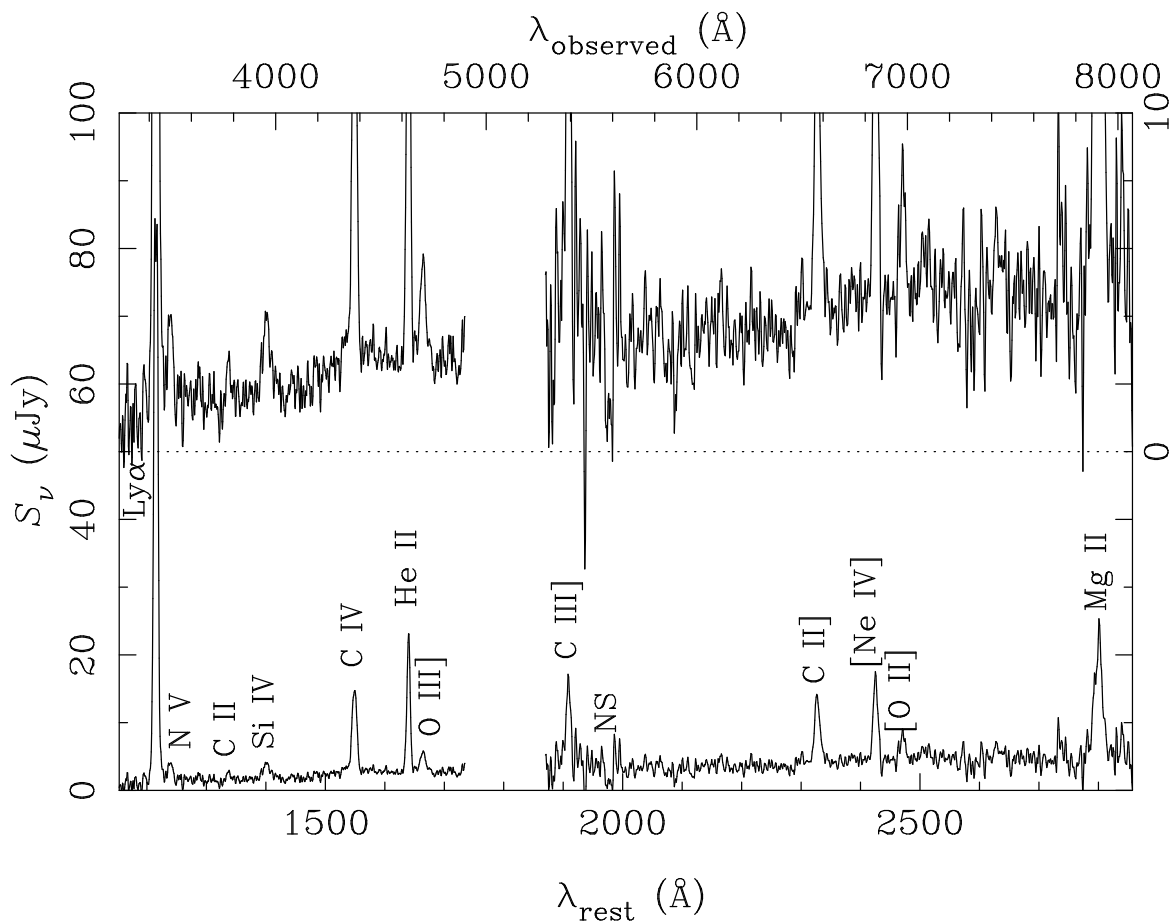}{10cm}{-90}{65}{65}{-245}{380}
\caption[]{Optical spectrum of 3C~256, taken with the Double
Spectrograph on the Hale 5-m telescope. The major emission lines are
labeled, as is the region strongly affected by night sky emission
(`NS'). The upper trace has had its $y$-scale expanded by a factor of
5 and shifted upwards by the amount indicated by the dotted line and
the scale on the right-hand axis.}
\label{fig:optspek}
\end{figure}

3C 256 was observed with the double spectrograph (Oke \& Gunn 1982) on
the Hale 5-m telescope at Palomar Observatory over the course of 5
nights in January 1990 and April 1991. A 2\arcsec-wide slit was used,
oriented along ${\rm PA} = 148\arcdeg$ (i.e., along the optical major
axis). The blue spectra were obtained with a 300\,lines\,mm$^{-1}$
grating blazed at 4000\,\AA\ and the red channel employed a
316\,lines\,mm$^{-1}$ grating blazed at 7500\,\AA\. Each exposure
lasted 3000--4000\,s and the object was recentered in the slit by
offsetting from a nearby star every 2 hours. The spectra were reduced
and calibrated in a standard manner. The flux calibration was based on
observations of Ross~374, Feige~67 and BD~+26\arcdeg2606. All of the
1-D spectra (extracted along a 5\farcs5 aperture) were combined, after
scaling those observed through cirrus (present on 2 of the 5 nights)
to match the spectra taken under photometric conditions. We present
the final, reduced spectrum in Fig.~\ref{fig:optspek}.

\subsection{Infrared spectroscopy}

A low-resolution near-infrared spectrum was obtained on the night of
1997 Jun 15 using the Near-Infrared Camera (NIRC; Matthews \& Soifer
1994) on the Keck I telescope. The 150 lines mm$^{-1}$ {\it JH\/}
grism was used with a 0\farcs7 wide slit, giving a resolution $R
\approx 80$. The slit was oriented at ${\rm PA} = 112\arcdeg$, which
was the closest to the major axis that could be achieved while still
being able to acquire a guide star. Individual 200\,s exposures were
obtained at five different positions along the slit, separated from
each other by 5\arcsec. The process was then repeated, to obtain a
total exposure time of 2000\,s. Each set of five exposures was
median-filtered and this median image subtracted from each individual
exposure to provide first-order sky subtraction. More accurate
subtraction of sky lines was performed on each exposure by fitting a
quadratic function of position to each column of the image, which was
then subtracted. The individual frames were registered and a spectrum
extracted through a 3\arcsec\ aperture.

\begin{figure}
\plotfiddle{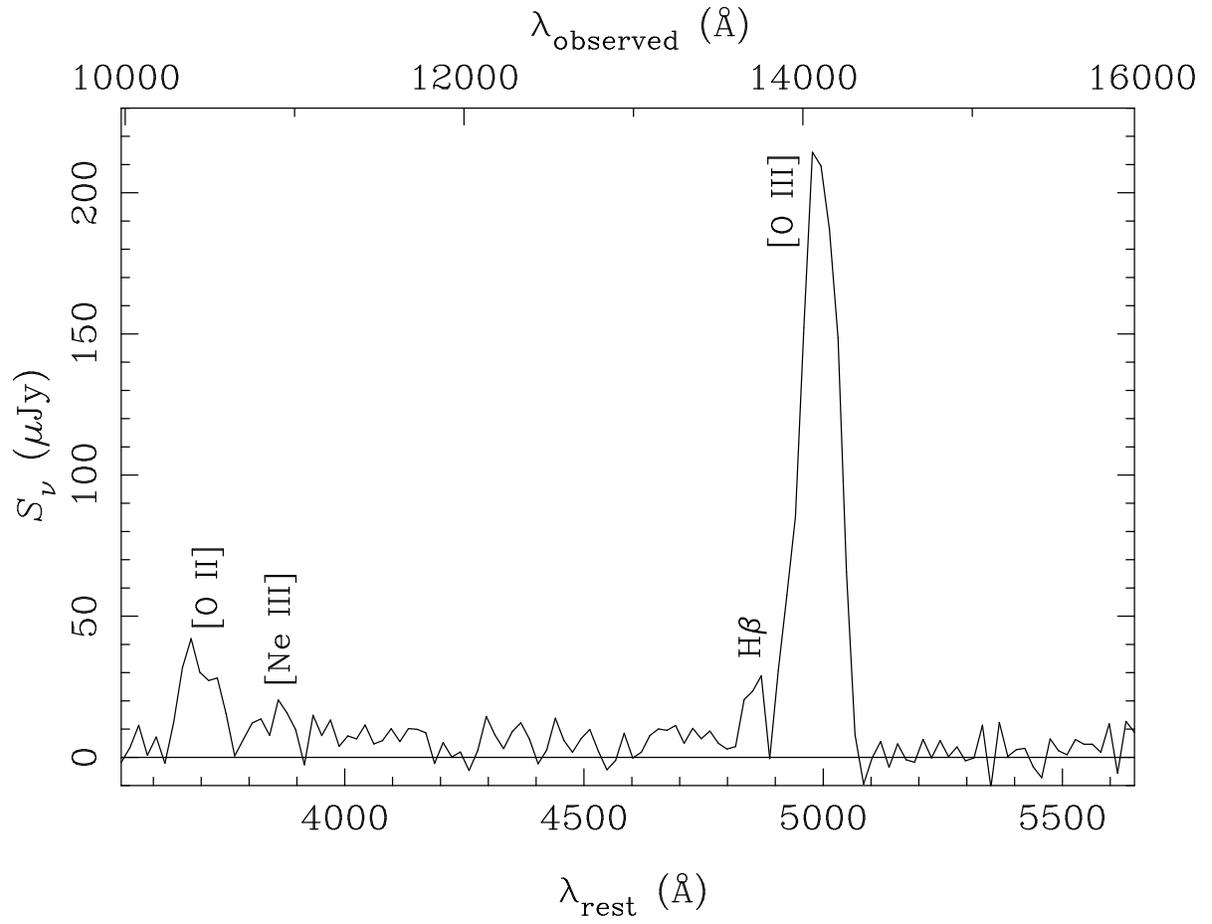}{10cm}{-90}{65}{65}{-245}{380}
\caption[]{Near-infrared spectrum of 3C~256, taken with NIRC on the
Keck I telescope. The major emission lines are labeled.}
\label{fig:irspek}
\end{figure}

Five exposures were also taken of the G5V star BD +32\arcdeg2290 and
reduced in a similar manner, and used to correct the 3C~256 spectrum
for atmospheric absorption and provide flux calibration. Short
exposure images in the $J$ and $K$ filters image were taken
immediately before and after the spectroscopy to enable an accurate
determination of the slit location. The reduced spectrum is presented
in Fig.~\ref{fig:irspek}.

\subsection{Optical imaging}

\begin{figure}
\plotone{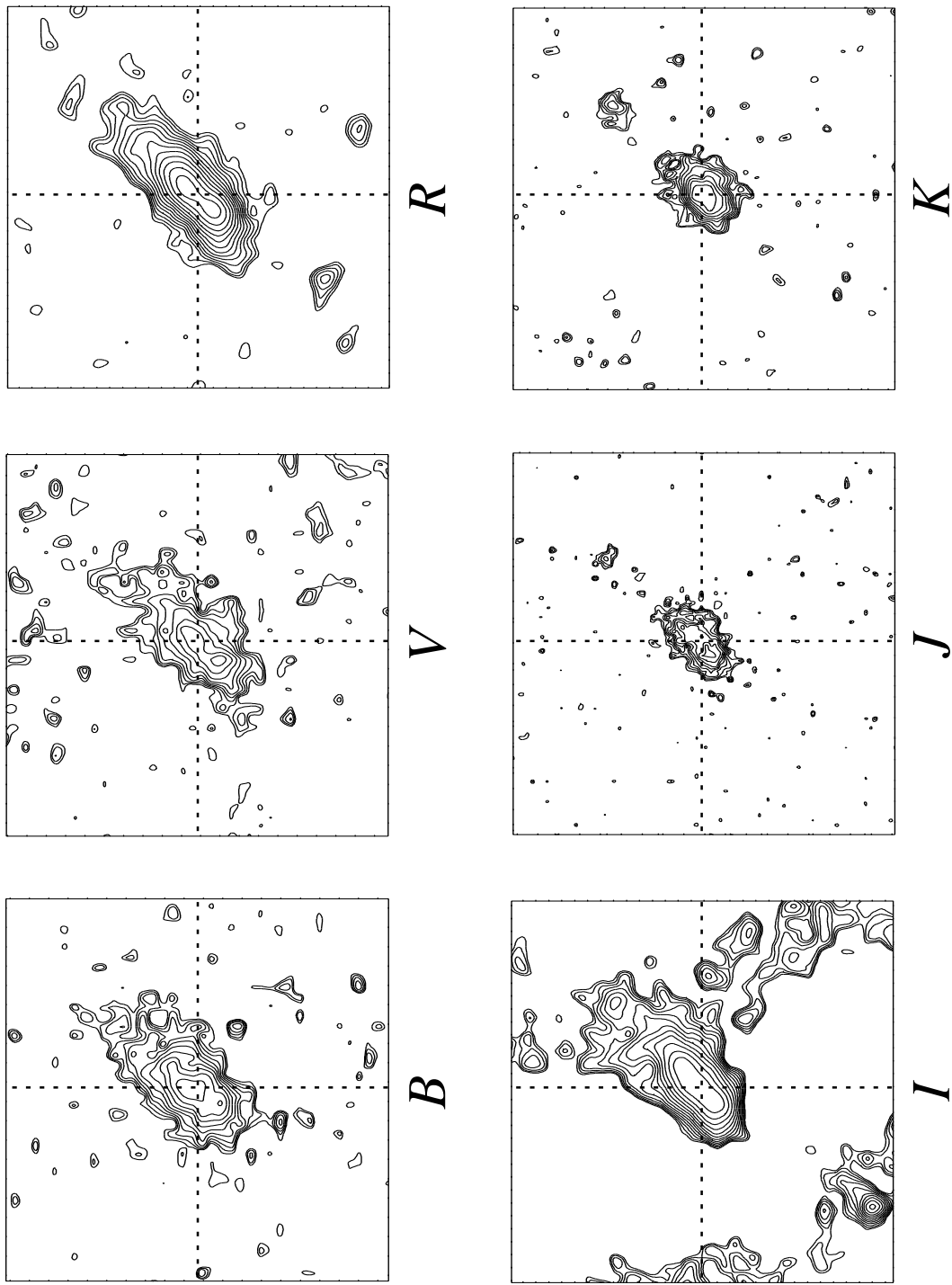}
\caption[]{Continuum images of the radio galaxy 3C~256 (the images marked
with a dagger in Table~\ref{tab:photom} are the ones shown). Each
image is 12\arcsec\ on a side with North up and East to the left, and
the dotted lines indicate the location of the peak of the $K$ emission
(the astrometric accuracy is $\sim 0\farcs2$).  Contour levels are
spaced at intervals of 0.25\,mag. The structure seen in background of
the $I$ image is due to fringing.}
\label{fig:cont}
\end{figure}

A list of the ground-based imaging observations used in this paper is
presented in Table~\ref{tab:photom}; a number of these have been
presented elsewhere (Spinrad \& Djorgovski 1984; Le F\`{e}vre et al.\
1988). In addition, an image was taken with the Mayall 4-m telescope
at Kitt Peak National Observatory though an interference filter with
central wavelength 8800\,\AA\ and FWHM 550\,\AA, and a $V$-band image
was also obtained at the Hale Telescope. Both sets of data were
reduced in a similar manner, using standard techniques. Flux
calibration was performed using observations of the field of M~92
(Christian et al.\ 1985; $V$-band image) or spectrophotometric
standards from the list of Massey \& Gronwall (1990; 8800\,\AA\
image). The best images taken through the $B$, $V$, $R$, and $I$
filters are shown in Fig.~\ref{fig:cont}.

\subsection{Infrared imaging}

A number of new near-infrared images were also obtained for this
paper, as listed in Table~\ref{tab:photom}. These include $J$ and
$K$-band images taken with NIRC on the Keck I telescope, $J$ and $K$
images taken with the infrared array camera IRIM (a $62 \times 58$
InSb array with a plate scale of 0\farcs39\,pixel$^{-1}$) on the
Mayall Telescope, and an $H$-band image taken using the Prime Focus
Infrared Camera (PFIRCAM; Jarrett et al.\ 1994) on the Hale
Telescope. Observations and subsequent reduction were performed
following standard `jittering' methods. Flux calibration was
determined from observations of photometric standards from the lists
of Elias et al.\ (1982; Mayall data), Persson et al.\ (1998; Keck
data), or Casali \& Hawarden (1992; Hale data). The $K$-band image
from the Mayall telescope is the coaddition of several mosaics taken
on the nights of 1989 Apr 17, 1991 May 24, and 1992 Mar 22. The $J$
and $K$ images from Keck are presented in Fig.~\ref{fig:cont}.

\subsection{HST imaging}

Initial {\it HST\/}   imaging  observations of 3C~256  were  made with
WF/PC in Cycle~1, but no useful and reliable morphological information
could be  extracted from them,  due  to a poor  signal-to-noise ratio,
lack of good flatfields,  and the complexity of the  pre-refurbishment
point spread function. These data are therefore  not discussed in this
paper.

\begin{figure}
\plotone{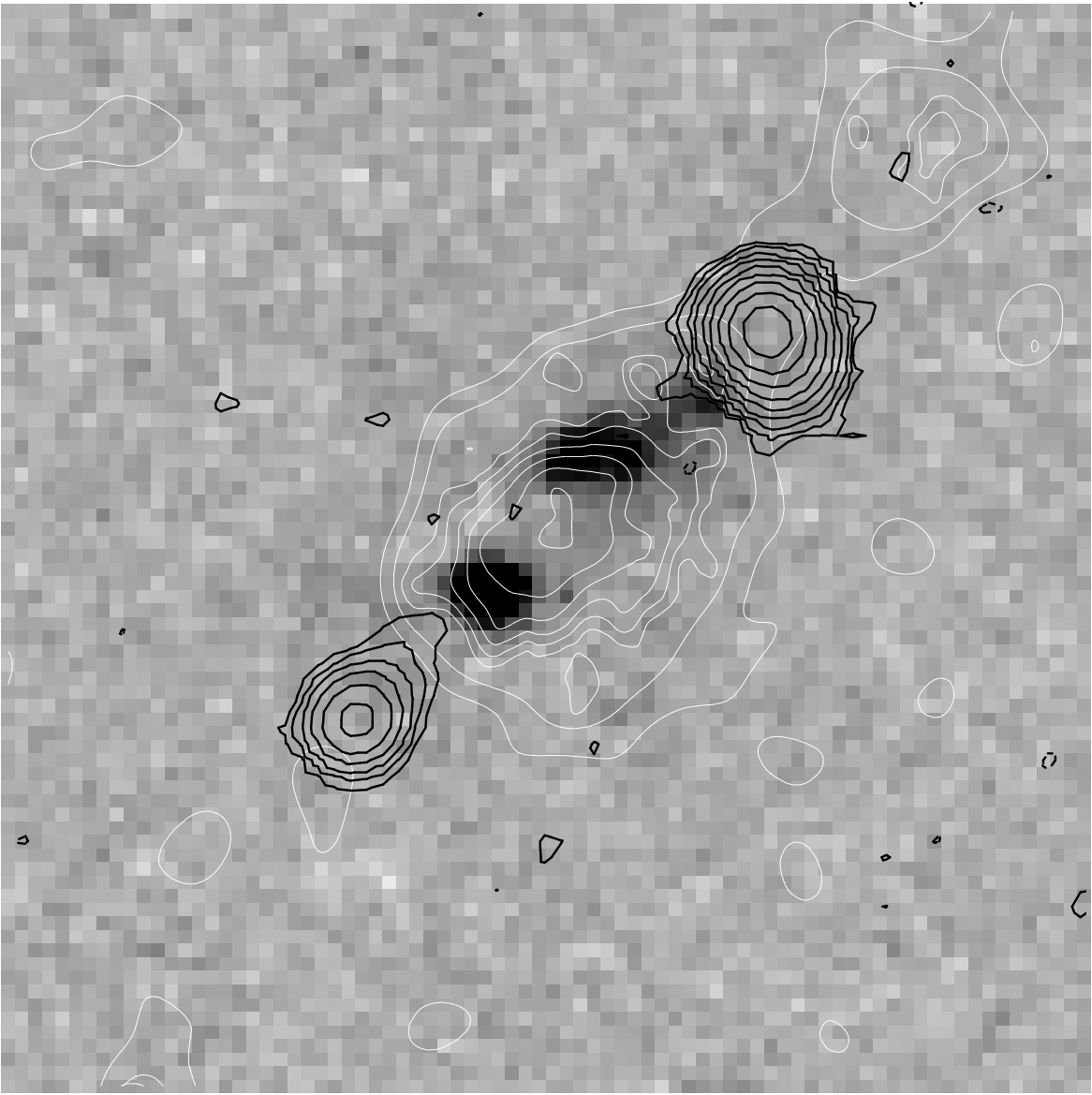}
\caption[]{{\it Greyscale: HST\/} F336W (Ly$\alpha$) image; the flux
does not drop to zero between the bright peaks. {\it White
contours:\/} Keck $K$-band image. The bottom two contours are from a
smoothed version of the image to accentuate the bridge of low surface
brightness emission connecting the radio galaxy to the NW object. {\it
Black contours:\/} VLA 5\,GHz map, reproduced from D96. The peak of
the $K$ emission lies between the two bright Ly$\alpha$ peaks; the
radio map has been registered assuming that the $K$ peak lies midway
between the radio lobes. The image is $8\arcsec \times 8\arcsec$, with
North up and East to the left.}
\label{fig:montage}
\end{figure}

The source was re-observed with {\it HST\/}/WFPC2 in the F336W filter
on 1996 May 22. Ly$\alpha$, at a wavelength of 3433\,\AA, lies close
to the peak transmission of this filter and, since the observed
equivalent width of the emission line is $\sim 1500$\,\AA\ and the
width of the filter is only 381\,\AA, it dominates the observed
flux. The source was placed on the WF3 chip, providing a scale of
0\farcs100\,pixel$^{-1}$. Eight orbits were spent on the source, for a
total of 21,400\,s of integration. The data were processed through the
STScI pipeline (Voit et al.\ 1997) and cosmic rays were rejected and
the data combined in the normal manner using the {\sc iraf} task {\it
crrej\/}. In Fig.~\ref{fig:montage} we show an overlay of the F336W
and $K$-band images and the 5\,GHz radio map of D96.

\section{Results}

\subsection{Spectroscopy}
\label{sec:spec}

We have used the {\sc specfit} software (Kriss 1994; available as part
of {\sc iraf/stsdas}) to measure line fluxes and equivalent widths
from our spectra. We fitted a Gaussian line profile and linear
continuum level to a 200\,\AA-wide region of the optical spectrum
(1000\,\AA-wide for the infrared spectrum), centered on the emission
line. In instances where two emission lines were separated by less
than 100\,\AA\ (500\,\AA), we fitted the continuum and two Gaussians
simultaneously to a region covering 100\,\AA\ (500\,\AA) on both sides
of each line. It was confirmed in all cases that the flux of the
best-fitting Gaussian was very close to that produced by a direct
measurement of the flux above the best-fitting continuum level. The
emission-line fluxes and rest-frame equivalent widths are presented in
Table~\ref{tab:lines}.

The power law which provides the best fit to our spectrum has a
spectral index $\alpha = 1.66 \pm 0.04$ ($S_\nu \propto
\nu^{-\alpha}$), very much redder that the $\alpha = 1.1 \pm 0.1$
found by D96. However, if we restrict our wavelength range to that
between the \ion{C}{4} and \ion{Mg}{2} lines, common to both spectra,
a much bluer spectral index ($\alpha = 1.29 \pm 0.13$) is found,
consistent with D96's value. This indicates a change in spectral shape
near 1500\,\AA\ (rest-frame). We return to this point in the next
section.

\subsection{Spectral energy distribution}
\label{sec:sed}

We co-registered the various images by computing a geometric
transformation based on the locations of stars in the images, and
constructed the spectral energy distribution (SED) of 3C~256 in the
optical--near-infrared (rest-frame ultraviolet--optical) by making
photometric measurements through a 4\arcsec\ aperture, centered on the
peak of the $K$-band emission. This aperture is large enough to
encompass all the high surface brightness emission from the galaxy,
and therefore errors introduced by poor registration of the images
will be smaller than the photometric uncertainties.

We correct the observed broad-band fluxes for the presence of strong
emission lines. We smooth the F336W image, which is dominated by
Ly$\alpha$, to the resolution of our ground-based images and compare
the fluxes in the spectroscopic and photometric
apertures\footnote{Recently-obtained {\it HST\/} F555W images confirm
that the ultraviolet continuum shares the Ly$\alpha$ morphology
(Jannuzi et al.\ 1999).}.  By assuming the other emission lines share
the Ly$\alpha$ morphology, we can thus use their spectroscopic fluxes
to estimate their contributions to the broad-band photometry. We
correct the optical photometry and $J$-band photometry using the
spectra of Figs~\ref{fig:optspek} and \ref{fig:irspek},
respectively. The $H$-band flux is corrected for the presence of the
[\ion{O}{1}] $\lambda\lambda$6300,6364 doublet assuming
[\ion{O}{1}]~$\lambda$6300/[\ion{O}{3}]~$\lambda5007 = 0.03$. No major
emission lines lie within the $K$-band. We estimate the uncertainty in
our aperture correction to be $\sim 10$\%, based on the variations
which result from different smoothing scales. However, this usually
translates to a $\sim 1$--2\% error in the line-corrected flux, much
less than the random photometric uncertainty. The exception is the
line-corrected flux from the F336W image ($1.8 \pm 0.7\,\mu$Jy), whose
large uncertainty is due to the high equivalent width of the
Ly$\alpha$ line.

\begin{figure}
\plotfiddle{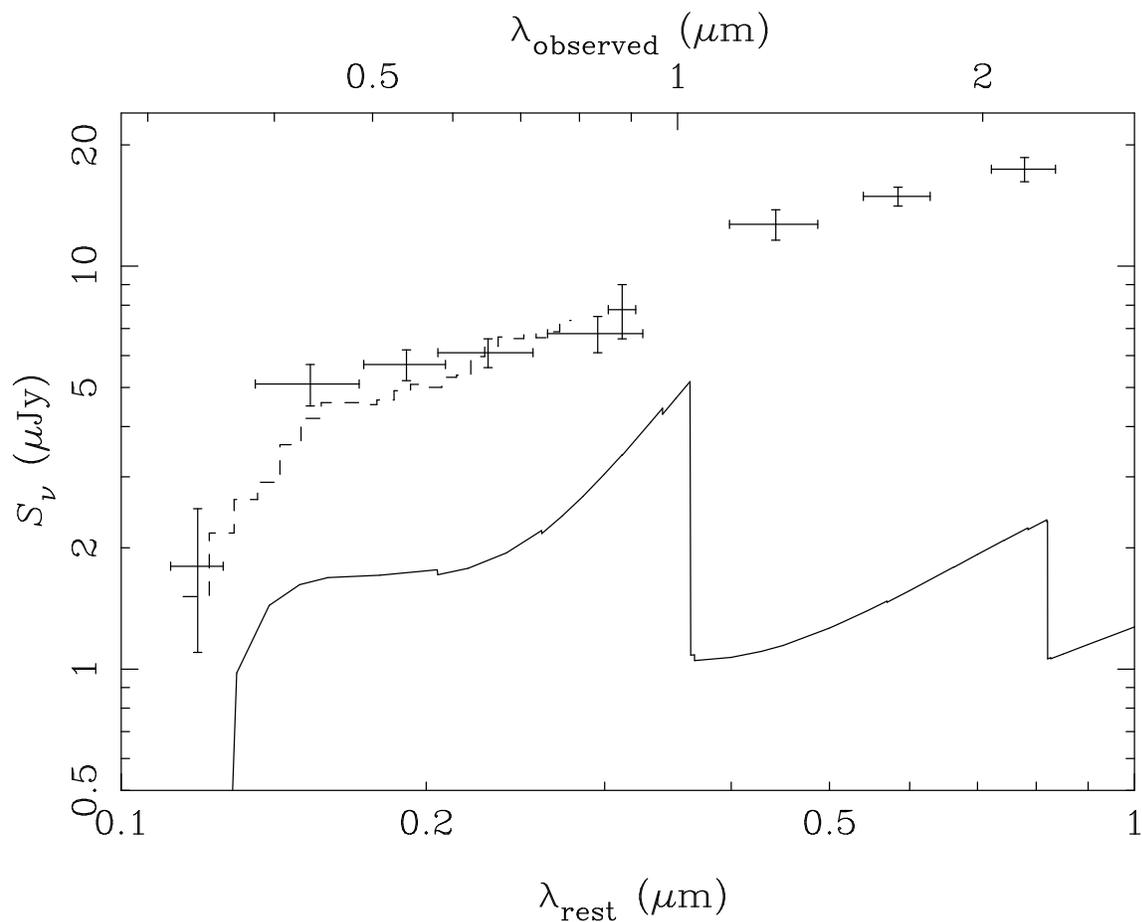}{10cm}{-90}{65}{65}{-245}{380}
\caption[]{Rest-frame spectral energy distribution of 3C~256. The
crosses indicate broad-band photometric measurements which have been
corrected for the presence of strong emission lines, and the dashed
line is the continuum from the Hale spectrum, rebinned to 200\,\AA\
resolution after removal of the emission lines, and scaled as
described in the text. The solid line is the nebular continuum
determined from the H$\beta$ flux. The flux density is plotted as
measured in the observed frame.}
\label{fig:sed}
\end{figure}

When constructing the SED, we usually take the weighted mean of fluxes
measured through similar filters.  However, there is a significant
difference between the $J$-band filters used for the Mayall and Keck
observations, in that the Keck filter extends to longer wavelengths,
and includes the strong [\ion{O}{3}]~$\lambda\lambda$4959,5007
doublet. Even though these lines lie in a region of poor and uncertain
atmospheric transmission, their equivalent width is so large that they
contribute significantly to the measured flux. We therefore use the
Mayall measurement of the $J$-band flux alone, whose bluer cutoff
excludes these strong lines.  We present the complete SED in
Fig.~\ref{fig:sed}.

The SED exhibits three major features which must be reproduced in any
viable model. First, the optical--infrared color is very blue ($V - K
= 2.9$); in fact 3C~256 is bluer even than a star forming dwarf Irr at
this redshift. Secondly, longward of 1500\,\AA\ (rest-frame), the SED
can be approximated by a power law with $\alpha \approx 1$. Finally,
there is a pronounced decrease in flux shortward of 1500\,\AA\ (seen
most obviously in the spectrum rather than the ${\rm F336W} - B$
continuum color), which we shall refer to as the ``UV rollover''.

\begin{figure}
\plotfiddle{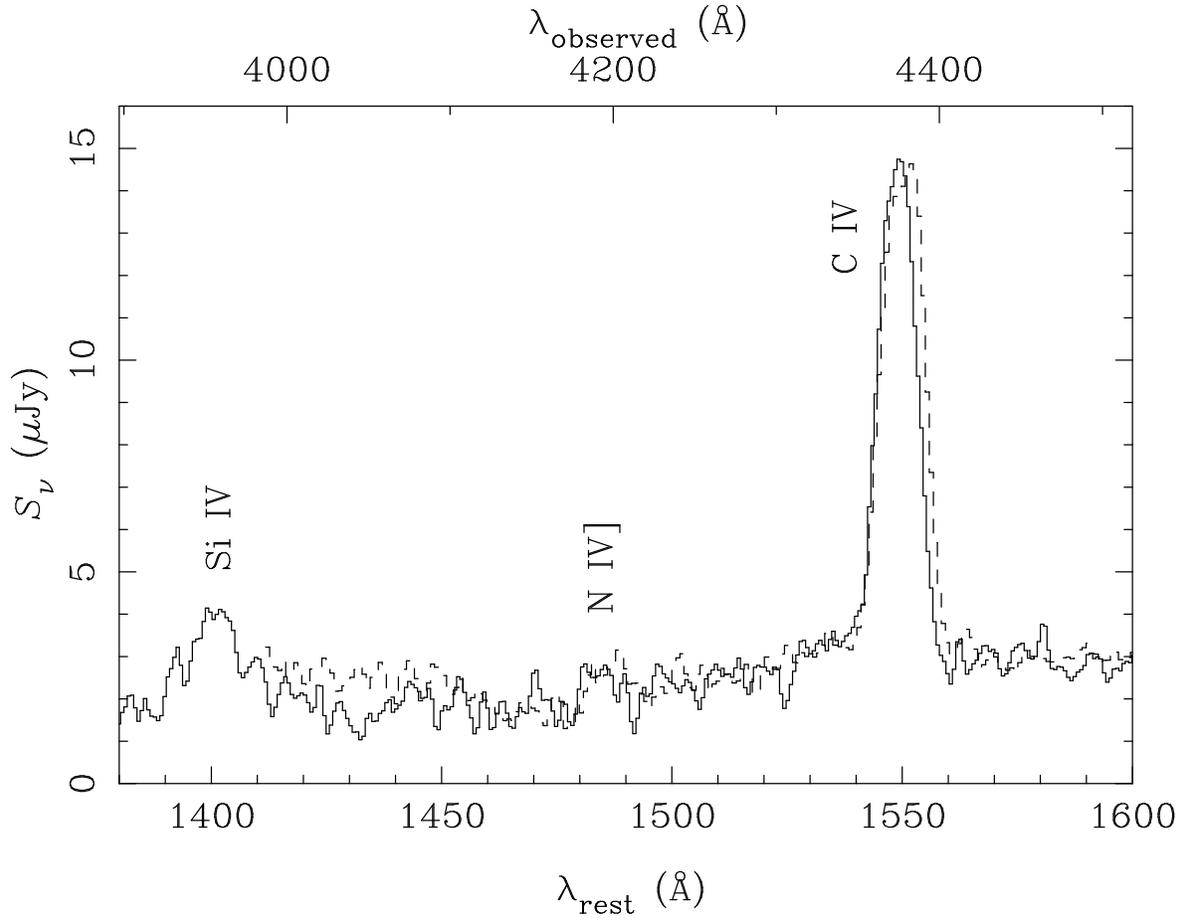}{10cm}{-90}{65}{65}{-245}{380}
\caption[]{Our spectrum (solid line) and that of D96 (dashed line, after
scaling) around the region of the UV rollover. Note that both spectra show
a decrease in flux density shortward of \ion{C}{4}, but the continuum level
increases again at shorter wavelengths in D96's spectrum.}
\label{fig:rollover}
\end{figure}

The importance of the UV rollover will become apparent later, and we
spend some time here confirming its reality, since it is not seen in
D96's spectrum. In Fig.~\ref{fig:rollover}, we plot our spectrum and
that of D96 (scaled to match ours redward of \ion{C}{4} since their
spectrum was taken in non-photometric conditions through a slightly
smaller aperture) in the wavelength region of interest. Both spectra
agree very well around the \ion{C}{4} line and show a decrease in flux
at $\lambda_{\rm rest} \approx 1480$\,\AA; however, D96's rises again
below $\sim 1450$\,\AA, and is $\sim 50$\% higher than ours shortward
of this wavelength. This rise produces a broad absorption feature
which D96 interpret to be a BAL cloud seen in the scattered light
spectrum.  We reject the possibility that our spectrum is incorrectly
flux calibrated, since the spectroscopic Ly$\alpha$ flux agrees well
with that measured from images.  The difference between the spectra
may be due to aperture effects (D96 used a $4\farcs1 \times 1\arcsec$
aperture, compared to our $5\farcs5 \times 2\arcsec$), but this does
not concern us since we are attempting to determine the overall SED,
as measured in a 4\arcsec\ aperture, rather than the detailed spatial
distribution of individual components. Importantly, the spectrum of
Spinrad et al.\ (1985) also shows the UV rollover, and rules out the
possibility of differential atmospheric refraction as the cause of the
rollover, since their observing technique positioned the major axis of
the entrance aperture along the parallactic angle.

As with D96, we note the absence of any strong absorption lines in the
spectrum, with the possible exception of a feature at 1720\,\AA\ (a
blend of \ion{N}{4}, \ion{Si}{4} and \ion{Al}{2}; e.g., Fanelli et
al.\ 1992), although this is a marginal detection at best. Our typical
signal-to-noise ratio of 4 per 2\,\AA\ pixel implies a rest-frame
equivalent width detection limit of $\sim 2$\,\AA\ if the absorption
lines are a few hundred km\,s$^{-1}$ in width.

\subsection{Morphology}

The {\it HST\/} F336W image reveals two strong emission peaks,
separated by 1\farcs3, which can be identified with components $a$ and
$b$ from Le F\`{e}vre et al.\ (1988). There is also a third peak,
further to the NW, which is not seen in Le F\`{e}vre et al.'s image
but might plausibly be the cause of the extension of $b$, and some
diffuse emission which contributes about 20\% of the total flux. If
the Ly$\alpha$ emission is produced by photoionization from a central
source, it should be located within two oppositely-directed cones,
whose common apex marks the location of the hidden active nucleus.
From Fig.~\ref{fig:montage}, we measure the half-opening angles of
such cones to be 28\arcdeg, and the projected axis of the cones to be
at ${\rm PA = 146\arcdeg}$, slightly offset from the radio axis
position angle of 132\arcdeg\ (D96).

Despite having rather poorer spatial resolution, our $B$ and $V$
images show a similar structure to Le F\`{e}vre et al.'s data. The
structure at $K$ is very different, however, with a central peak
appearing between the Ly$\alpha$-emitting regions. The color of the
central region of 3C~256 is $R - K = 3.4$, while the more extended
emission has $R - K \approx 2.7$. This difference should be considered
a lower limit to the true color variation, as seeing effects will
smear the more strongly-peaked $K$-band image, causing some
redistribution of flux from the center to the outer regions.

In addition, there is an object located 4\arcsec\ NW of 3C~256 along
the radio axis. There appears to be low surface brightness emission
connecting the companion to the radio galaxy, visible in both the $J$
and $K$ images, and so we speculate that the two are physically
connected. The companion is only convincingly seen in the
near-infrared images and, while there may be a marginal detection in
the deep CFHT $I$-band image, the fringing precludes any photometry.

\section{Analysis}

Our goal in this paper is to disentangle the various components which
contribute to the overall spectral energy distribution of 3C~256 by
using our new imaging and spectroscopy data, together with published
polarimetry. Although it would clearly be helpful to model distinct
regions of the source separately, the F336W image indicates that
significant structure occurs on a scale of about 0\farcs5. Our
seeing-limited ground-based observations therefore preclude such an
analysis, and instead we investigate the integrated SED of 3C~256
through a 4\arcsec\ aperture. All the fluxes used in this section are
either measured directly through such an aperture, or have been scaled
from the optical spectrum by a factor of 1.2, to account for the
difference in aperture sizes.

\subsection{Scattered light}
\label{sec:scattered}

The polarimetric properties of 3C~256 all suggest that the polarized
radiation is produced by the scattering of emission from a central
source (Jannuzi et al.\ 1995; D96), presumably non-stellar emission
from the active nucleus. Jannuzi et al.\ had no clear preference for
electrons or dust as the scattering particles, while D96 favored
electrons. In addition to the arguments advanced by D96, the spectral
index of the polarized flux ($\alpha = 1.1 \pm 0.1$; D96) is redder
than the mean for radio-loud quasars (e.g., Baker \& Hunstead 1995;
Willott et al.\ 1998), which argues against the substantial bluening
that dust scattering often produces. Also, theoretical models of dust
scattering predict a fairly sharp change in polarization at $\lambda
\lesssim 2000$\,\AA\ (Kartje 1995; Manzini \& di Serego Alighieri
1996), which is not seen in 3C~256 (D96). We therefore assume electron
scattering throughout the remainder of this paper. Since this
mechanism is achromatic, the scattered light spectrum will be the same
as the spectrum of the central AGN seen by the scatterers (i.e., at
angles close to the radio axis). We model this with the core-dominated
quasar spectrum of Baker \& Hunstead (1995), artificially ``reddened''
by 0.6 powers of $\nu$ to match the observed spectral index of the
polarized flux (D96), and extrapolated as a power law longward of
H$\alpha$.

We can estimate the fraction of scattered light by considering the
lack of prominent broad lines in our spectrum. The broad permitted
lines from the quasar nucleus should be scattered into our line of
sight together with the non-stellar continuum. We fit the spectral
region around \ion{C}{4}~$\lambda$1549 with narrow and broad Gaussians
and determine a $3\sigma$ upper limit of 25\,\AA\ to the rest-frame
equivalent width of any broad line. Like the limits D96 found for
\ion{Mg}{2}~$\lambda$2800, this is substantially lower than the mean
value for the equivalent width of \ion{C}{4} in radio-loud quasars
(e.g., Steidel \& Sargent 1991; Baker \& Hunstead 1995), but within
the observed range. The absence of a broad emission line in our
spectrum is therefore suggestive of the presence of non-scattered
flux, but is certainly not conclusive. The lower signal-to-noise ratio
in our spectrum at Ly$\alpha$ means that even a broad line with an
equivalent width of 100\,\AA\ (cf., a mean of 70\,\AA; Baker \&
Hunstead 1995) is consistent with the data.

\subsection{Nebular continuum}
\label{sec:nebular}

Broadly speaking, the overall SED is consistent with an $\alpha
\approx 1$ power law. Since this is also the spectral shape of the
polarized flux, it appears possible to ascribe the total flux spectrum
entirely to scattered light with a constant fractional polarization of
$\sim 11$\%. However, Dickson et al.\ (1995) have shown that nebular
continuum emission can be very important in powerful radio galaxies,
and it is essential that we consider its contribution to the
ultraviolet flux of 3C~256.

\begin{figure}
\plotfiddle{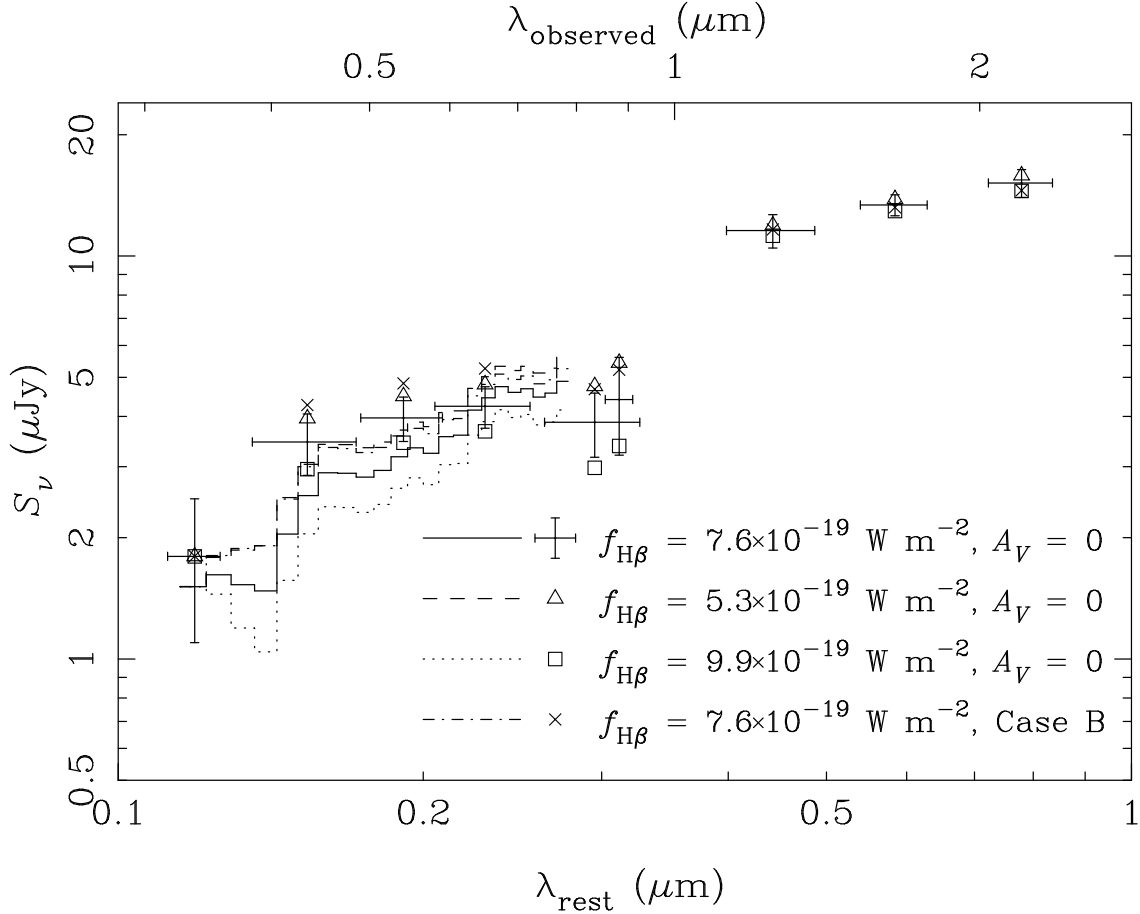}{10cm}{-90}{65}{65}{-245}{380}
\caption[]{The rest-frame spectral energy distribution of 3C~256 (as
Fig.~\ref{fig:sed}), after removal of the nebular continuum emission,
according to different assumptions. The error bars and solid line
represent the subtraction of unreddened nebular continuum emission
corresponding to the observed H$\beta$ flux. The triangles and dashed
line, and squares and dotted line are for different values of $f_{\rm
H\beta}$, representing the $\sim 30$\% uncertainty in the observed
flux. The crosses and dash-dot line are the result if the
H$\beta$/Ly$\alpha$ ratio is assumed to be intrinsically equal to the
low-density Case~B recombination value, and deviates from it due to a
foreground screen of dust obeying Pei's (1992) SMC extinction law.}
\label{fig:neb}
\end{figure}

After applying an aperture correction, we determine the H$\beta$ flux
in our 4\arcsec\ aperture to be $(7.6 \pm 2.3) \times
10^{-19}$\,W\,m$^{-2}$. This implies Ly$\alpha$/H$\beta \approx 9$, a
value typical of high-redshift radio galaxies (e.g., McCarthy, Elston,
\& Eisenhardt 1992). We assume $N_{\rm He}/N_{\rm H} = 0.1$ and
that 5\% of the helium is in the form of \ion{He}{2}, based on the
inferred ratio \ion{He}{2}~$\lambda$1640/H$\beta \approx 1$ and the
recombination coefficients of Osterbrock (1989), although the result
is rather insensitive to this value. Using the tabulated emission
coefficients of Aller (1987) for $T_{\rm e} = 10^4$\,K, the nebular
continuum emission is shown by the solid curve in Fig.~\ref{fig:sed}.
In order to see more clearly the significance of the nebular emission
to the overall SED, in Fig.~\ref{fig:neb} we plot the SED with the
nebular emission (assumed to be unreddened) removed. The shape is
dramatically different, with a clear break appearing at $\lambda_{\rm
rest} \sim 4000$\,\AA, which was previously obscured by the hydrogen
recombination continuum.

Since this break is the most important evidence in favor of a
substantial stellar population in 3C~256, we investigate how its
strength is affected by different assumptions about the nebular
continuum emission. Given the uncertainty in the H$\beta$ flux, we
construct alternative SEDs where the nebular continuum has been
reduced or increased by one-third (although note that the high
observed [\ion{O}{3}]/H$\beta = 22$ ratio argues against our H$\beta$
flux being too high by a large amount). We also produce an SED
where we have assumed that the observed ratio of Ly$\alpha$/H$\beta
\approx 9$ deviates from the low-density Case~B recombination value of
23 (e.g., Ferland \& Osterbrock 1985) due to a foreground dust screen,
which also affects the observed nebular continuum. We use Pei's (1992)
Small Magellanic Cloud extinction law (the reason for this choice will
become apparent later), although nearly identical results are obtained
if we use Pei's Galactic law or Calzetti's (1997) empirical ``recipe''
which is based on observations of starburst galaxies. The implied
reddening is $E(B-V) = 0.13$. All three alternative SEDs are also
plotted in Fig.~\ref{fig:neb}, where it can be seen that the variation
caused by these different assumptions is comparable to the photometric
uncertainties. Furthermore, the Balmer jump persists throughout all
the different assumptions, as shown in Fig.~\ref{fig:neb}. We
therefore adopt our original assumptions throughout the remainder of
this analysis.

\subsection{A young stellar population}

The pronounced jump in the continuum level near 4000\,\AA\ implies
starlight is a significant contributor to the total SED, both above
and below this jump. Any stellar population must be fairly young
(identifying this jump as the Balmer jump, not the 4000\,\AA\ break),
since neither the 2600 nor 2900\,\AA\ breaks displayed by more evolved
stellar populations are observed in our spectrum (see also D96), and
the continua on both sides of the break are quite blue. However, three
features of the revised SED are inconsistent with a population of
arbitrarily young (hot) stars, namely the strong Balmer jump,
relatively flat UV continuum and the UV rollover. All three of these
are typical of main sequence stars of late B or early A type (e.g.,
Fanelli et al.\ 1992). Such stars only dominate the UV flux after the
$\sim 100$\,Myr needed for hotter stars to complete their evolution.

While no stellar absorption features are detected in our spectrum, the
measured upper limit of 2\,\AA\ is typical of the narrow blends seen
in A-type main sequence stars (Fanelli et al.\ 1992). This is before
considering the effects of dilution from the scattered and nebular
components, and the masking of absorption features by the strong
emission lines. The observations are therefore not at odds with the
notion that stars contribute significantly to the ultraviolet flux.

We compare our observations with synthetic stellar spectra produced
with the GISSEL96 code (Bruzual \& Charlot 1993, 1999), adopting a
Salpeter (1955) IMF which extends from 0.1--125\,$M_\odot$. The upper
mass cutoff is a fairly critical parameter in determining the age of
the stellar population, since it is possible to reach the same main
sequence turnoff much more rapidly than 100\,Myr if more massive stars
are never formed. On the other hand, the lower mass cutoff is
important only in determining the total mass of stars.

\begin{figure}
\plotfiddle{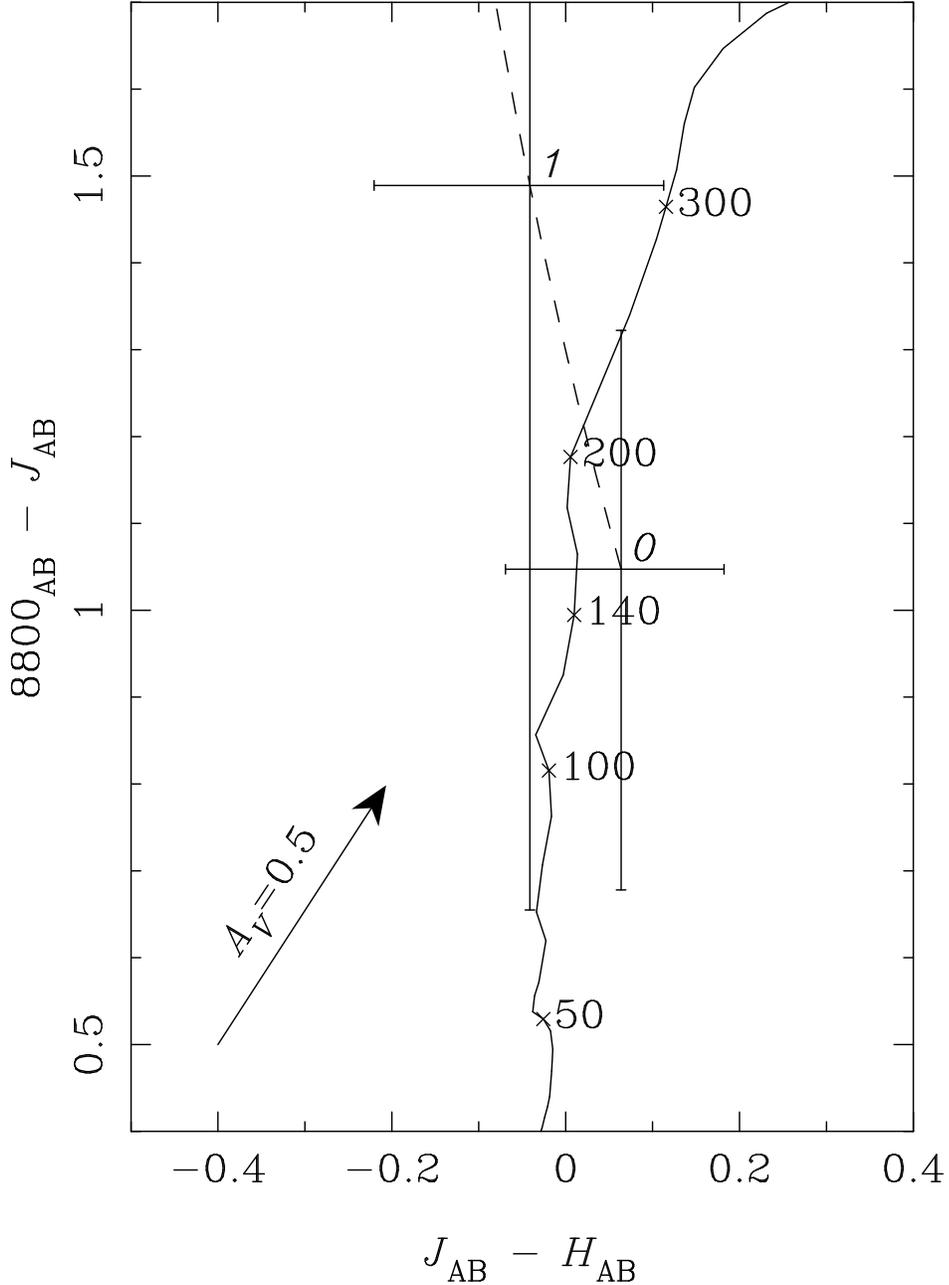}{18cm}{0}{75}{75}{-200}{0}
\caption[]{Fitting a quasar spectrum and starlight to the 8800{\it
JH\/} photometry. The fluxes are expressed as AB magnitudes, where
$m_{\rm AB} = {\rm constant} - 2.5 \log S_\nu$. The solid line gives
the colors of a stellar population formed in an instantaneous burst,
with the numbers listing the age in Myr. The dashed line represents
the photometry (corrected for nebular emission) after the removal of
the quasar spectrum described in the text. The error bars are plotted
whenever the strength of the scattered component in the observed $V$
band is an integral number of $\mu$Jy, as indicated by the italic
numerals. The effect of 0.5\,mag of reddening (the SMC law of Pei
(1992) is adopted) on the observed colors is shown by the arrow.}
\label{fig:break}
\end{figure}

We estimate the age of the young stellar population using the observed
$8800-J$ and $J-H$ colors, after correcting for the nebular continuum
and scattered emission. In Fig.~\ref{fig:break}, these colors are
compared with those of a stellar population formed in an instantaneous
burst and seen at $z=1.824$. The age of the stellar population is
estimated to be between 70 and 400\,Myr for an instantaneous burst
(effectively independent of the assumed metallicity), with a
best-fitting value of 220\,Myr. It is clear from Fig.~\ref{fig:break}
that the age of the stellar population is constrained almost entirely
by the $8800-J$ color (i.e., the strength of the Balmer jump) and,
since this wavelength baseline is short, the results are fairly
insensitive to the spectral shape of the scattered continuum. Even if
the scattered radiation does deviate significantly from our adopted
spectrum, a young stellar population is still required to fit the SED
of 3C~256. This analysis also provides an estimate of the strength of
the scattered component, and our conclusion that $\sim 20$\% of the
observed optical continuum flux is provided by this component is in
line with the estimate of D96. Approximately half the rest-frame
ultraviolet continuum therefore comes from starlight.

Although we assume an instantaneous burst, the uncertainty in this age
gives some indication of the possible duration over which star
formation could have occurred. Models where the star formation rate
(SFR) follows an exponential decline cannot provide a good fit to the
overall SED unless the characteristic time scale is small ($\lesssim
20$\,Myr), since the UV rollover demands that most of the 1500\,\AA\
flux arises from stars older than about 100\,Myr, and hence the
present star formation rate must be low. Models where the SFR is
constant for a finite period of time and then drops to zero are more
successful, with there being little difference in the SEDs of a
200\,Myr-old population whose stars formed instantaneously, and one
in which they formed over a period of 50\,Myr.

\subsection{The red core}

All three components we have so far discussed are too blue to explain
the red color at the center of 3C~256, which is also associated with a
change in the galaxy's morphology. We therefore need to address the
possibility of a fourth component.

The red central color and strong $K$-band peak can be simply explained
by a compact red component superimposed on a spatially flat blue
continuum. If we assume that the spatial color variations in this blue
continuum are small (supported by the similarities between the F336W
and F555W {\it HST\/} images), we can estimate the flux of the
putative core component by scaling the $R$ image to match the extended
flux in the $K$ image and subtracting it. The residual source has a
flux of $\sim 3\,\mu$Jy, and is slightly resolved compared to the
point spread function of a star in the image.  The core cannot
therefore be due to a lightly reddened quasar nucleus, as has been
seen in other 3C radio galaxies (Simpson, Rawlings, \& Lacy 1999, and
references therein; see \S\ref{sec:unify} for the implications of this
result), and is presumably stellar in origin.

The core could either be intrinsically red, and dominated by old
stars, or be reddened by dust. The presence of dust is suggested by
the Ly$\alpha$ morphology, which has a brightness minimum at the
location of the core. Rest-frame ultraviolet continuum images of other
distant radio galaxies show similar morphologies, and Dickinson, Dey,
\& Spinrad (1995) have argued that the $z=1.206$ host galaxy of the
radio source 3C~324 must suffer substantial reddening ($A_V \gtrsim
1$) for it not to be seen in their WFPC2 image. On the other hand, if
the companion object is associated with 3C~256, its red colors support
an intrinsically red stellar population. This source is redder than
3C~256 by $0.38 \pm 0.11$\,mag in $J-K$ and more than 1.2\,mag in
$R-K$ (after correcting for the presence of emission lines in the
radio galaxy).

We somewhat arbitrarily prefer the dust-lane model to explain the red
color of the central regions of 3C~256. In attempting to model the
SED, we assume that the underlying population has the same age as the
blue population of the previous section. Since the observed morphology
of 3C~256 indicates that this red component does not contribute
significant flux shortward of the $K$-band, its nature and spectral
shape have little impact on our determination of the strengths of the
other components.

\subsection{The four component model}

\begin{figure}
\plotfiddle{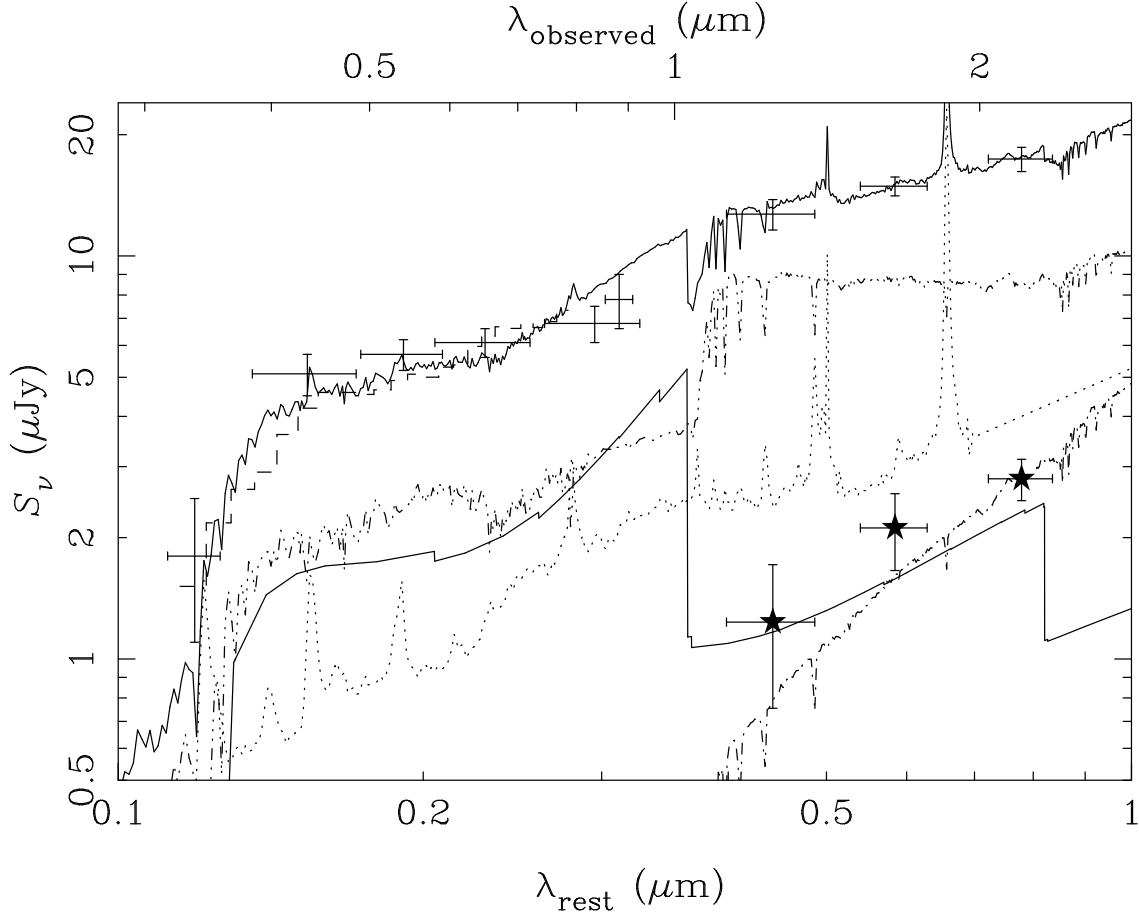}{10cm}{-90}{65}{65}{-245}{380}
\caption[]{The SED of Fig.~\ref{fig:sed} with the best fitting
four-component model. The upper solid line shows the total flux from
nebular continuum emission (lower solid line), an electron-scattered
quasar spectrum (dotted line), reddened (by $A_V = 2$\,mag) stars
(dash-dot line), and unreddened stars (dash-dot-dot-dot line). The
stellar population is a 180\,Myr old instantaneous burst of mass $1.3
\times 10^{11}\,M_\odot$, with the mass approximately equally divided
between the reddened and unreddened components. The filled stars with
error bars indicate the flux of the companion object (measured in a
2\arcsec\ aperture), scaled so that its $K$-band flux is equal to that
of the reddened stellar component.}
\label{fig:fit}
\end{figure}

Since we have been able to infer a substantial amount of information
about the various components already, it is not difficult to find the
four-component model which provides the best fit to the observations.
The difficulty in objectively weighting the spectroscopic and
photometric data forces us to determine the best fit by eye. However,
we have also performed a grid search to minimize the $\chi^2$
quality-of-fit statistic between the broad-band photometry and the
model, and confirmed that this was not significantly better than our
adopted fit. The characteristics of this fit are summarized in
Table~\ref{tab:fit}, and the fit is shown graphically in
Fig.~\ref{fig:fit}.

We determine a mass of $6.6 \times 10^{10}\,M_\odot$ for the UV-bright
stellar component. This either requires a peak SFR of $\gtrsim
3000\,M_\odot$\,yr$^{-1}$ for exponentially declining star formation,
or a constant SFR of $\sim 1300\,M_\odot$\,yr$^{-1}$ for a period of
$\sim 50$\,Myr. Although high, we note for comparison that Dey et al.\
(1997) estimate that the SFR could be as high as
$1100\,M_\odot$\,yr$^{-1}$ in the $z = 3.80$ radio galaxy 4C~41.17.

\subsection{The three component model}

Although we argued above for a fourth component to explain the red
core, we investigate whether a reddened, centrally-peaked, young
stellar population can explain the structure of 3C~256. The dust
responsible for this reddening would need to have an extinction law
which is steep in the far UV (1200--1600\,\AA) and devoid of a
2200\,\AA\ bump. This rules out a Galactic extinction law, since the
strength of the bump correlates with the steepness of the law in the
far UV (Cardelli et al.\ 1989). We therefore consider the extinction
law of the Small Magellanic Cloud (SMC) and adopt the parametrized SMC
extinction law of Pei (1992).

The UV rollover and $\alpha \approx 1$ power law can be reproduced
with an $\alpha \approx 0$ power law seen through $A_V \approx
0.5$\,mag of extinction. The intrinsically bluer color obviously
requires a younger stellar population than determined in the previous
section, and Fig.~\ref{fig:break} indicates that foreground reddening
also lowers the age of the stellar population needed to fit the $8800
- J$ color to $\sim 120$\,Myr. This scenario obviates the need for a
fourth component to explain the red core, since the color of the
reddened stars is $R - K = 3.2$, consistent with that observed in the
central regions of the galaxy.

\begin{figure}
\plotfiddle{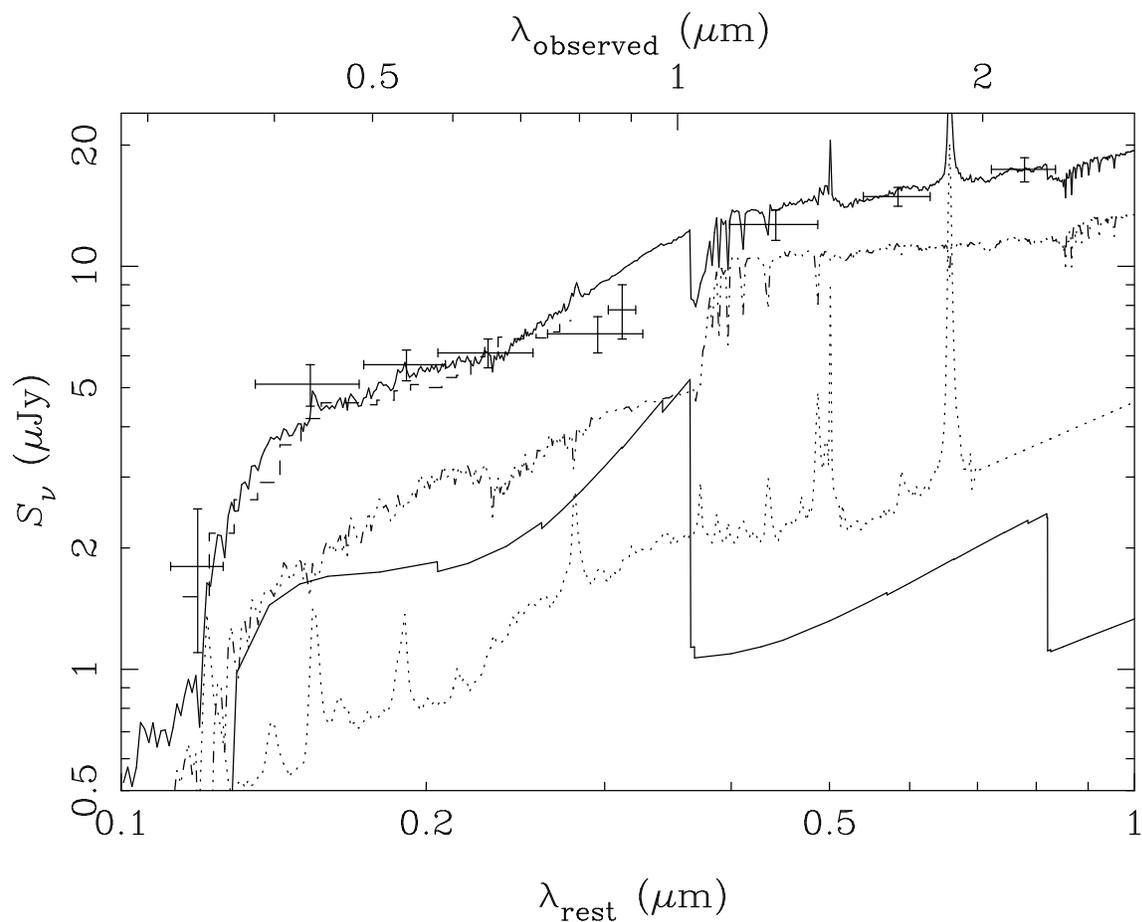}{10cm}{-90}{65}{65}{-245}{380}
\caption[]{The SED of Fig.~\ref{fig:sed} with the best fitting
three-component model. The upper solid line shows the total flux from
nebular continuum emission (lower solid line), an electron-scattered
quasar spectrum (dotted line), and a 130\,Myr-old stellar population
of mass $8.2 \times 10^{10}\,M_\odot$, reddened by $A_V = 0.25$\,mag
(dash-dot-dot-dot line).}
\label{fig:fit2}
\end{figure}

We find the best-fitting three-component model, again adopting a
by-eye fit and confirming that it is not significantly worse than a
$\chi^2$-minimization fit to the broad-band photometry. The results of
this fit are presented in Table~\ref{tab:fit} and
Fig.~\ref{fig:fit2}. Our present data do not favor either model over
the other, although it is hoped that {\it HST\/} imaging and
polarimetry will enable more detailed modeling and a conclusion to be
drawn. However, since both models require a substantial young stellar
population, our conclusion that one is present in 3C~256 is not
dependent on which is correct, but rather on the accuracy of our
H$\beta$ flux measurement and the observed UV rollover.

\section{Discussion}

\subsection{The alignment effect in 3C~256}
\label{sec:align}

In both of our models, half of the observed $V$-band light arises from
either scattered light or nebular continuum. In schemes which attempt
to unify extragalactic radio sources, both of these should both be
aligned with the radio source, since the rest-frame ultraviolet
radiation escapes in two oppositely-directed cones along the radio
axis. We have observational evidence to support this, since both the
polarized continuum (which traces the scattered quasar light) and the
line emission (which traces the nebular continuum) are extended along
the radio axis. The optical-radio alignment seen in 3C~256 is
therefore naturally explained. Although the fraction of the total
light at $K$ produced by the scattered and nebular emission is lower,
at 35\%, it is still high enough to produce a pronounced near-infrared
alignment effect, due to the absence of a strong, unaligned red
component such as an evolved stellar population.

While the aligned optical light is provided in approximately equal
amounts by the nebular continuum and scattered light, the latter
component produces the majority of the aligned $K$-band light. In
fact, the scattered quasar light provides a larger fraction of the
total light at $K$ than at $V$, and therefore the fractional $K$-band
polarization should be larger, if the intrinsic polarization of the
scattered light remains constant. Simply scaling by the fractional
contributions from the scattered component, we predict that the
$K$-band polarization should be about 16\% (in a 4\arcsec\ aperture),
although the exact value is somewhat dependent on the specific
geometry.

Even without the scattered emission, the nebular emission could
produce a strong optical alignment by itself: it would contribute over
40\% of the total $V$-band flux in the absence of the scattered
light. This alignment should be most pronounced at an observed
wavelength of $\sim 1$\,\micron, where the Balmer continuum is
strongest. Unfortunately, our 8800\,\AA\ image has insufficient
signal-to-noise to make a quantitative statement about the alignment
strength, but it should be possible to confirm spectroscopically the
importance of nebular emission in 3C~256. Blanketing from high-order
Balmer lines causes the effective wavelength of the Balmer jump from
the young stellar population to be longer than the Balmer limit at
3646\,\AA. As both Figs~\ref{fig:fit} and \ref{fig:fit2} show, this
results in a deep (the observed equivalent width is $\sim 140$\,\AA),
broad absorption feature around 3700\,\AA, which should be detectable
with moderate-resolution spectroscopy ($R \sim 300$) even at a fairly
low signal-to-noise ratio.

What then of the jet-induced star formation scenario? On a cursory
inspection, this appeared quite favorable for 3C~256, due to the
similar extents of the radio and optical emission, and the
near-infrared alignment. However, the age of the stellar population,
at 100--200\,Myr is older than any plausible estimate for the age of
the radio source, as we now demonstrate.

The separation of the radio lobes is 4\farcs3, which means that the
hotspots have propagated an average distance of $24 (\sin
\theta)^{-1}$\,kpc, where $\theta$ is the angle between the radio axis
and the line of sight. This angle is likely to be large because no
radio core is seen in D96's radio map with a limit $\log R < -3.36$
($R$ is the ratio of core to extended radio luminosity at $\nu_{\rm
rest} = 5$\,GHz; we use D96's spectral indices for the lobes and
assume a flat core spectrum to determine the $K$-corrections). This is
much lower than typical values of $\log R \approx -2.5$ (e.g.,
Morganti et al.\ 1997; Simpson 1998), ruling out the Doppler boosting
which occurs when the radio axis is close to the line of sight (e.g.,
Orr \& Browne 1982). Estimates of the hotspot advance speed in
powerful radio galaxies are typically $v_{\rm hs}
\approx 0.03$--0.15$c$ (Liu, Pooley, \& Riley 1992; Scheuer 1995),
although the speed may be $\sim 0.01c$ in young sources still within
their host galaxies (e.g., Fanti \& Fanti 1994). A realistic age for
the radio source is therefore
\[
t \sim 1.6 \times 10^6 (v_{\rm hs}/0.05c)^{-1} (\sin
\theta)^{-1}\,{\rm yr},
\]
which is about two orders of magnitude lower than the age we have
determined for the stellar population.

Of course, the `age' of the stellar population is, in fact, merely a
measure of the location of the main sequence turnoff, which the UV
rollover and Balmer jump strength constrain to be late B/early A-type.
The true age could therefore be much shorter if the IMF is biased
against high-mass stars. Such a suggestion is, however, counter to
claims made on theoretical grounds that the IMF should be biased
against {\em low\/}-mass stars (Larson 1977). In addition, the
spectrum of 4C~41.17, the most convincing case for jet-induced star
formation, clearly shows features associated with hot, young stars
(Dey et al.\ 1997) which are strong enough to have been detectable in
our spectrum. The discrepancy in the ages of the stellar population
and the radio source therefore demands that the trigger for the most
recent episode of star formation must have come from elsewhere. It can
be shown, following Scheuer \& Williams (1968), that 100\,Myr is
sufficient time for relic lobes to have dissipated through synchrotron
losses, thereby allowing the possibility of multiple episodes of radio
source activity (e.g., Roettiger et al.\ 1994). On the other hand,
Daly (1990,1992b) has shown that galactic rotation will disrupt any
radio--optical alignment over such a timescale, and so this scenario
appears unlikely.

One question we have not answered is whether the young stellar
population in 3C~256 is aligned with the radio source. This is
impossible to determine with images of different depths and
resolutions, especially when the contribution from the aligned
components is $\gtrsim 25$\%, even in the least-affected ({\it J\/})
band. Also, the presence of a dust lane perpendicular to the radio
source, as we have postulated might be the cause of the red core,
would produce an alignment even when the underlying population is not
aligned. Multicolor images at $\sim 0\farcs1$ resolution might help us
to disentangle the structures of the various components and address
this question.

\subsection{The stellar mass of 3C~256}

The fact that 3C~256 is underluminous in the $K$--$z$ diagram (it is
more than a magnitude fainter than the locus of other 3C radio
galaxies; e.g., Eales et al.\ 1997) has been used to infer a
protogalactic nature for the source, where it has yet to form the bulk
of its eventual stellar mass. Our analysis lends further weight to
this picture.

Best, Longair, \& R\"{o}ttgering (1998) find that the
optical--infrared SEDs of 3CR radio galaxies can be fairly
well-modeled by a flat-spectrum aligned component and an old stellar
population which provides nearly all of the $K$-band light. On the
other hand, about 30\% of the $K$-band light in 3C~256 is non-stellar,
and at least half of the remainder originates from young stars whose
mass-to-light ratio is much lower than that of more evolved
populations. Although we cannot rule out the presence of an additional
stellar component which is so heavily reddened as to contribute very
little flux, there is no evidence to support its existence in 3C~256,
or its absence in other radio galaxies. It therefore appears that the
stellar mass of 3C~256 is less than that of similar radio galaxies by
a factor of a few.

We attempt to quantify this by calculating the stellar mass of our
model SED, and comparing it with the masses derived by Best et al.\
(1998). Since these authors also used the Bruzual \& Charlot spectral
synthesis code, there will be no systematic differences introduced.
For the four-component model, the mass of the young, blue stellar
component is fairly well-determined ($6.6 \times 10^{10}\,M_\odot$),
but the same is not true for the red component which is responsible
for the near-infrared core. We modeled it earlier as a reddened
version of the same young stellar population, and derived a similar
mass as for the unreddened component, but this mass depends on the
extinction, which is poorly constrained. In addition, the lack of
spectral information means that it could instead be a population of
older stars, with a higher mass-to-light ratio. The best-fitting
unreddened ``old'' population is 1--1.5\,Gyr old and has a mass of
$\sim 5$--$6 \times 10^{10}\,M_\odot$, slightly dependent on the
duration over which the star formation took place. The total stellar
mass is therefore very similar for both models of the red component.
For the three-component model, the mass of the single stellar
component is $8.2 \times 10^{10}\,M_\odot$.

These are lower limits to the total stellar mass for each model, since
there may be additional starlight, either heavily reddened or from an
old stellar population, which contributes very little visible light.
For a short burst of star formation at $z = \infty$, a mass of $7
\times 10^{10}\,M_\odot$ is required to produce 1\,$\mu$Jy of
(observed-frame) $K$-band light. This mass increases by 70\% for every
magnitude of foreground visual extinction. On balance we doubt that
the total stellar mass is likely to be greater than $2 \times
10^{11}\,M_\odot$, much less than the $7 \times 10^{11}\,M_\odot$
which is the median stellar mass for the lower-redshift 3C sample of
Best et al.\ (1998; we modify their results to account for the
difference in our adopted cosmologies), and more than a factor of two
lower than any of their estimates.

Although our uncertainty as to the nature of the red core means we
cannot confirm the putative protogalactic nature of 3C~256, the low
inferred stellar mass compared to other radio galaxies does suggest
that it is not yet mature. Since the observed polarization requires a
substantial mass of gas (Jannuzi et al.\ 1995; D96), and even the gas
mass needed to produce the observed line emission is large ($6 \times
10^{10}\,M_\odot$ for $n_{\rm e} = 5$\,cm$^{-3}$), it is quite
conceivable that 3C~256 could undergo further bursts of star
formation. Alternatively, it could acquire more mass through mergers.
The companion object may be a galaxy which is in the early stages of a
merger with 3C~256. Although spectroscopic confirmation that the
companion is at the same redshift as 3C~256 may not be feasible (it
has $R \gtrsim 25$), the red optical--infrared colors ($R - J > 2.0$)
suggest the presence of a spectral break around 1\,\micron, supporting
a similar redshift. It is worth noting that the near-infrared colors
of the companion are very similar to those of the red core (see
Fig.~\ref{fig:fit}), and may therefore be a coeval stellar population,
if the red core is old, rather than reddened.

\subsection{Implications for unification scenarios}
\label{sec:unify}

We briefly discuss what implications the absence of an unresolved
near-infrared core has for unification schemes. The two quasars in the
3CR-based catalogue of Laing, Riley \& Longair (1983) with redshifts
closest to 3C~256 both have $V \approx 18.4$. Assuming that 3C~256
would be similarly bright if its quasar nucleus were seen directly, a
power law spectrum with $\alpha \approx 1$ extended into the
near-infrared implies a $K$-band flux of 0.6\,mJy for the nucleus
alone. We therefore require $A_V \gtrsim 12$\,mag of rest-frame
extinction to produce a core with a flux below 1\,$\mu$Jy.  Most $z
\sim 1$ radio galaxies have nuclear obscurations in excess of 15\,mag
(Simpson et al.\ 1999), and therefore the extinction required to hide
the quasar nucleus from direct view at $K$ is not unreasonably large.

In both of our models, approximately one-fifth of the rest-frame UV
light is non-stellar radiation scattered into our line of sight. We
should therefore expect to see broad lines whose equivalent widths are
one-fifth of their intrinsic values. The limits we determined in
\S\ref{sec:scattered} are consistent with this fraction, even if the
intrinsic equivalent widths are significantly higher than the average
values for radio-loud quasars. A normal quasar-like central engine is
therefore not ruled out by their absence.

\section{Summary}

We have modeled the spectral energy distribution of the $z=1.824$
radio galaxy 3C~256. Although the overall SED is consistent with a
single power law, a clear break appears at $\lambda_{\rm rest}
\sim 4000$\,\AA\ after subtraction of nebular continuum emission which
is constrained by our infrared spectroscopy. Although no stellar
absorption lines are seen in our spectrum, this break indicates that
starlight is a major contributor to the overall SED. Our model
includes starlight, nebular continuum emission, and scattered AGN
light, and provides a good fit to the data. The dominant stellar
population has an age of 100--200\,Myr, and was formed in a fairly
short burst with a peak star formation rate of $\gtrsim
1000\,M_\odot$\,yr$^{-1}$.

In our model, the alignment effect seen in 3C~256 is due to a
combination of nebular emission and scattered light, with the latter
dominating at longer wavelengths. The pronounced near-infrared
alignment is due to the absence of a bright evolved stellar
population. Simple arguments based on size show that the radio source
must be younger than 100\,Myr and therefore is unlikely to have
influenced the most recent burst of star formation. We predict that
3C~256 should have substantial near-infrared polarization, estimating
a value of about 16\% in the $K$-band. We also expect it to show a
strong absorption feature (the observed-frame equivalent width is
140\,\AA\ in our model) comprised of the Balmer edge at 3646\,\AA\ in
emission, and blanketing from high order Balmer lines in the young
stellar population.

We have estimated the total stellar mass of 3C~256 to be no more than
$2 \times 10^{11}\,M_\odot$, rather lower than estimates for other 3C
radio galaxies. Arguments based on the polarization properties suggest
that there is a large mass of gas associated with 3C~256, and
therefore further major bursts of star formation are possible.

The specific case of 3C~256 presented here indicates the complex
nature of powerful radio galaxies and the importance of obtaining data
spanning as large a wavelength baseline as possible. In a future
paper, we shall present the results of {\it HST\/} imaging polarimetry
in the F555W filter, which should help us to understand more about
this interesting source.

\acknowledgments

This work has been supported by the National Aeronautics and Space
Administration, through {\it HST\/} grants GO-2698 and GO-5925 from
the Space Telescope Science Institute. B.~T.~J. acknowledges support
for his research from the National Science Foundation through their
cooperative agreement with AURA, Inc.\ for the operation of the
National Optical Astronomy Observatories. The W.~M. Keck Observatory
is operated as a scientific partnership among the California Institute
of Technology, the University of California and the National
Aeronautics and Space Administration. The Observatory was made
possible by the generous financial support of the W.M. Keck
Foundation. Parts of this work were performed at the Jet Propulsion
Laboratory, California Institute of Technology under a contract with
NASA. We thank Arjun Dey and Olivier Le F\`{e}vre for supplying us
with some of the data used in this paper. C.~S. wishes to thank Clive
Tadhunter for an invaluable discussion which helped to direct this
work.

\clearpage

\clearpage

\begin{table}
\caption[]{Broad-band observations and fluxes measured in a 4\arcsec\
aperture.}
\label{tab:photom}
\begin{tabular}{cccrcrl@{ }l}
\tableline \tableline
\footnotesize
Filter & Telescope & Date\tablenotemark{a} & Exposure (s) &
Seeing\tablenotemark{b} & Flux ($\mu$Jy) & \multicolumn{2}{c}{Line
contamination ($\mu$Jy)\tablenotemark{c}} \\
\tableline
$B^\dagger$ & Mayall & SD84 & & 1\farcs3 & $6.3 \pm 0.6$ & 1.2 &
(\ion{Si}{4}, \ion{C}{4}, \ion{He}{2}, \ion{O}{3}])
\\
$V^\dagger$ & Hale & 1997 Jun 3 & $900\;\;$ & 1\farcs0 & $6.2 \pm 0.4$
& 0.6 & (\ion{C}{3}])
\\
$V$ & Mayall & SD84 & & 1\farcs2 & $6.6 \pm 0.7$ & 0.6 & (\ion{C}{3}])
\\
$R$ & Mayall & SD84 & & 1\farcs5 & $7.1 \pm 0.8$ & 0.5 &
(\ion{C}{2}], [\ion{Ne}{4}])
\\
$R^\dagger$ & CFHT & L88 & $3600\;\;$ & 0\farcs7 & $6.5 \pm 0.3$ & 0.6
& (\ion{C}{2}], [\ion{Ne}{4}])
\\
$I^\dagger$ & CFHT & L88 & $2700\;\;$ & 0\farcs9 & $7.4 \pm 0.7$ & 0.6
& (\ion{Mg}{2})
\\
8800 & Mayall & 1992 May 1 & $1600\;\;$ & 1\farcs1 & $7.8 \pm 1.2$ &
0.0 &
\\
$J$ & Mayall & 1992 Mar 20 & $7200\;\;$ & 0\farcs8 & $12.8 \pm 0.7$ &
0.1 & (H$\gamma$)
\\
$J^\dagger$ & Keck I & 1998 Jan 18 & $900\;\;$ & 0\farcs6 & $13.2 \pm
0.7$ & 2.6\tablenotemark{d} & ([\ion{O}{3}])
\\
$H$ & Hale & 1998 Mar 25 & $1728\;\;$ & 1\farcs5 & $14.9 \pm 0.8$ &
1.1 & ([\ion{O}{1}])
\\
$K$ & Mayall & 1989--1992 & $15600\;\;$ & 0\farcs9 & $17.5 \pm 1.8$ &
0.0 &
\\
$K^\dagger$ & Keck I & 1994 Apr 5 & $1080\;\;$ & 0\farcs8 & $17.4 \pm
1.7$ & 0.0 &
\\
\tableline
\end{tabular}
\tablenotetext{a}{Images whose dates are listed as ``SD84'' were
presented in Spinrad \& Djorgovski (1984), and those listed as ``L88''
were used in Le F\`{e}vre et al.\ (1988).}
\tablenotetext{b}{The seeing is measured from the FWHM of stars in the
images.}
\tablenotetext{c}{The estimated contamination to the broad-band flux
from emission lines in the 4\arcsec\ photometric aperture is listed.
The observed fluxes listed in the previous column have not been
corrected for line contamination.}
\tablenotetext{d}{The line contribution is strongly dependent on the
atmospheric transmission at 1.4\,\micron\ at the time of the
observation, and is thus very uncertain.}
\tablecomments{Images marked $^\dagger$ are presented in
Fig.~\ref{fig:cont}.}
\end{table}

\clearpage

\begin{table}
\caption[]{Observed emission lines fluxes and rest-frame equivalent
widths.}
\label{tab:lines}
\begin{tabular}{lr@{}l@{$\,\pm\,$}r@{}lr@{}l@{}l}
\tableline \tableline
Line & \multicolumn{4}{c}{$f$ ($10^{-20}$\,W\,m$^{-2}$)} &
\multicolumn{3}{c}{EW (\AA)} \\
\tableline
\multicolumn{7}{c}{Optical fluxes ($5\farcs5 \times 2\arcsec$
aperture)} \\
\tableline
Ly$\alpha$                    & 542&   & 7&   & 501&  &$^{+91}_{-68}$ \\
\ion{N}{5}~$\lambda$1240      &  14&   & 9&   &  13&.1&$^{+11.8}_{-8.9}$ \\
\ion{C}{2}~$\lambda$1335      &   6&.2 & 0&.9 &   7&.0&$^{+1.3}_{-1.2}$ \\
\ion{Si}{4}~$\lambda$1400     &  21&.5 & 1&.9 &  23&.7&$^{+2.8}_{-2.6}$ \\
\ion{C}{4}~$\lambda$1549      &  52&.3 & 0&.8 &  38&.5&$\pm$1.1 \\
\ion{He}{2}~$\lambda$1640     &  54&.7 & 0&.5 &  43&.7&$^{+11.1}_{-7.5}$ \\
\ion{O}{3}]~$\lambda$1663     &  14&.6 & 1&.1 &  11&.0&$^{+4.0}_{-2.7}$ \\
\ion{C}{3}]~$\lambda$1909     &  42&.8 & 3&.3 &  27&.3&$^{+5.2}_{-4.5}$ \\
\ion{C}{2}]~$\lambda$2326     &  20&.3 & 0&.6 &  24&.1&$\pm$1.0 \\
{}[\ion{Ne}{4}]~$\lambda$2423 &  21&.2 & 0&.6 &  26&.1&$^{+1.3}_{-1.2}$ \\
{}[\ion{O}{2}]~$\lambda$2470  &   6&.7 & 0&.8 &   8&.4&$^{+1.2}_{-1.1}$ \\
\ion{Mg}{2}~$\lambda$2798     &  37&.7 & 2&.1 &  52&.7&$^{+6.5}_{-5.7}$ \\
\tableline
\multicolumn{7}{c}{Infrared fluxes ($3\arcsec \times 0\farcs7$
aperture)} \\
\tableline
{}[\ion{O}{2}]~$\lambda\lambda$3726,3729
& 154 & & 44& &  769&&$^{+829}_{-372}$ \\
{}[\ion{Ne}{3}]~$\lambda$3869 &  54&.7 &25&.3 & 273&&$^{+372}_{-167}$ \\
H$\beta$                      &  40&.7 &12&.6 & 156&&$^{+92}_{-60}$ \\
{}[\ion{O}{3}]~$\lambda\lambda$4959,5007
& 898 & & 40& & 3440&&$^{+1064}_{-706}$ \\
\tableline
\end{tabular}
\end{table}

\clearpage

\begin{table}
\caption[]{Summary of the properties of the best-fitting SED models.}
\label{tab:fit}
\begin{tabular}{lr@{.}lrrr}
\tableline \tableline
Component & \multicolumn{2}{c}{$f_V$ ($\mu$Jy)} & $f_V$ (\%) & $f_K$
($\mu$Jy) & $f_K$ (\%) \\
\tableline
& \multicolumn{5}{c}{Four-component model} \\
\tableline
Nebular continuum                 &    1&8  &    35\% & 2.1 & 12\% \\
Scattered quasar                  &    0&9  &    17\% & 4.0 & 23\% \\
Blue stellar pop\tablenotemark{a} &    2&5  &    48\% & 8.6 & 49\% \\
Red stellar pop\tablenotemark{b}  & $<0$&01 &  $<1$\% & 2.8 & 16\% \\
\tableline
& \multicolumn{5}{c}{Three-component model} \\
\tableline
Nebular continuum                   &    1&8  & 35\% &  2.1 & 12\% \\
Scattered quasar                    &    0&8  & 15\% &  3.5 & 20\% \\
Stellar population\tablenotemark{c} &    2&6  & 50\% & 11.7 & 68\% \\
\tableline
\end{tabular}
\tablenotetext{a}{A 180\,Myr-old population of mass $6.6 \times
10^{10}\,M_\odot$}
\tablenotetext{b}{A 180\,Myr-old population of mass $6.2 \times
10^{10}\,M_\odot$, reddened by $A_V = 2$\,mag}
\tablenotetext{c}{A 130\,Myr-old population of mass $8.2 \times
10^{10}\,M_\odot$, reddened by $A_V = 0.25$\,mag}
\end{table}

\end{document}